\documentclass{jltp}%
\usepackage{graphicx}
\usepackage{amsmath}
\usepackage{amsfonts}
\usepackage{amssymb}%
\title{Oxygen Phonon Branches in Detwinned YBa$_{\text{2}}$Cu$_{\text{3}}%
$O$_{\text{7}}$}
\author{D. Reznik$^{1,2}$, L. Pintschovius$^{1}$, W. Reichardt$^{1}$, Y. Endoh$^{3}$,
\\H. Hiraka$^{3}$, J. M. Tranquada$^{4}$,  S. Tajima$^{5}$, H.
Uchiyama$^{5}$, and T. Masui$^{5}$}

\address{$^{1}$Forschungszentrum Karlsruhe,
Institut f\"{u}r Festk\"{o}rperphysik,\\ Postfach 3640, D-76021
Karlsruhe, Germany\\$^{2}$Laboratoire Leon Brillouin,
C.E.A./C.N.R.S., F-91191-Gif-sur-Yvette \\CEDEX,
France\\$^{3}$Institute for Material Research, Tohoku University,
Katahira, Aoba-ku,\\ Sendai, 9808577, Japan\\$^{4}$Physics
Department, Brookhaven National Laboratory,\\ Upton, NY
11973\\$^{5}$Superconductivity Research Laboratory, ISTEC,
Shinonome, Koutu-ku, \\Tokyo, 135-0062, Japan }

\runninghead{D.Reznik et al.}{Oxygen Phonon Branches in Detwinned
YBa$_{2}$Cu$_{3}$O$ _{7}$}

\begin{document} \maketitle

\begin{abstract}
We report results of inelastic neutron scattering measurements of phonon
dispersions on a detwinned sample of YBa$_{\text{2}}$Cu$_{\text{3}}%
$O$_{\text{7}}$ and compare them with model calculations. Plane
oxygen bond stretching phonon branches disperse steeply downwards
from the zone center in both the \textbf{a} and the \textbf{b}
direction indicating a strong electron-phonon coupling. Half way
to the zone boundary, the phonon peaks become ill-defined but we
see no need to invoke unit cell doubling or charge stripe
formation: lattice dynamical shell model calculations predict such
behavior as a result of branch anticrossings. There were no
observable superconductivity-related temperature effects on
selected plane oxygen bond stretching modes measured on a twinned
sample.


\end{abstract}

Charge degrees of freedom associated with lattice vibrations
and/or distortions may be important to our understanding of many
physical properties of cuprate superconductors.\cite{1}
Longitudinal vibrations of planar oxygens soften strongly with
doping away from the zone center in all cuprates investigated so
far\cite{2,3}, which has been interpreted as a signature of strong
electron-phonon coupling. Up to now twinning has made it difficult
to obtain a clear experimental picture of this behavior in the
middle of the zone, where phonon dispersions are steep. To clear
up twinning-related ambiguities, we measured phonon dispersions by
inelastic neutron scattering on an untwined sample of
YBa$_{\text{2}}$Cu$_{\text{3}}$O$_{\text{7}}$. High-resolution
measurements on a large twinned sample confirmed these results.

\begin{figure}
[ptb]
\begin{center}
\includegraphics[
width=\textwidth
]%
{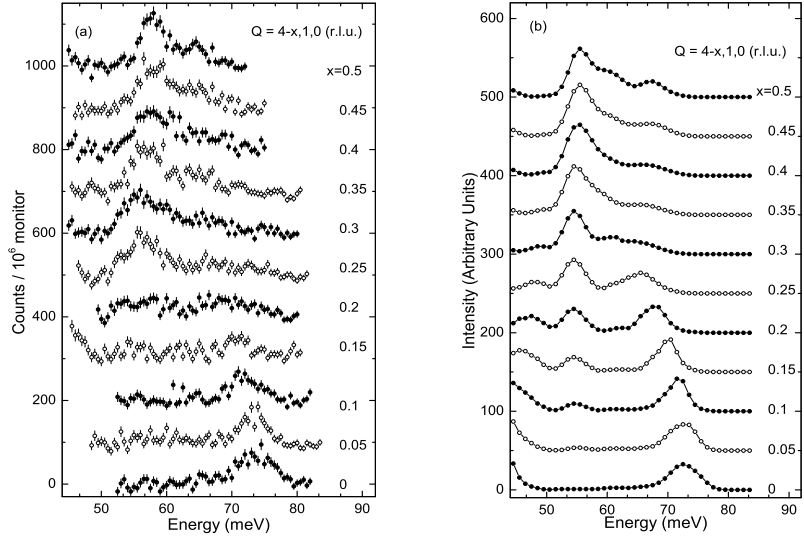}%
\caption{Measured (a) and resolution-corrected calculated (b)
neutron scattering intensity at \textbf{Q} = (4-X,1,0) r.l.u.
Background, linear in energy and independent of \textbf{Q}, has
been subtracted from
the experimental data. }%
\end{center}
\end{figure}
%
The untwinned sample consisted of close to 30 detwinned single
crystals with the mosaic spread of 3$^{\circ}$ and the total
volume of 0.8cm$^{3}$. The twinned sample consisted of 3 single
crystals of combined volume of 1.5cm$^{3}$ with the mosaic spread
of 2.2$^{\circ}$. The experiments were performed on the
triple-axis spectrometer 1T at the Orphee reactor using doubly
focusing monochromator (Cu111 and Cu220) and analyzer (PG002)
crystals. \textbf{a}-axis and \textbf{b}-axis atomic vibrations of
the $\Delta$1 symmetry were measured in the Brillouin zone
adjacent to wavevectors, \textbf{Q}, of (4,1,0) and (-1,4,0)
respectively. The data were fit with gaussian lineshapes on top of
a background assumed to be linear in energy, $\omega,$ and
independent of \textbf{Q}.

The starting point for our calculations was the common interaction
potential model\cite{4}, which is quite successful in describing
the phonon dispersions of a number of cuprates.
YBa$_{\text{2}}$Cu$_{\text{3}}$O$_{\text{7}}$ is treated as an
ionic compound and the interatomic interactions are modelled as a
sum of Coulomb forces and short range repulsive forces. The
polarizability of the atoms is treated using the shell model
formalism. For a metallic compound like
YBa$_{\text{2}}$Cu$_{\text{3}}$O$_{\text{7}}$ a term accounting
for screening by free carriers is added. This model gives a decent
description of the available data, which was further improved by
tuning the parameters. Further tuning was done to account for the
new data.

\begin{figure}
[ptb]
\begin{center}
\includegraphics[
width=\textwidth
]%
{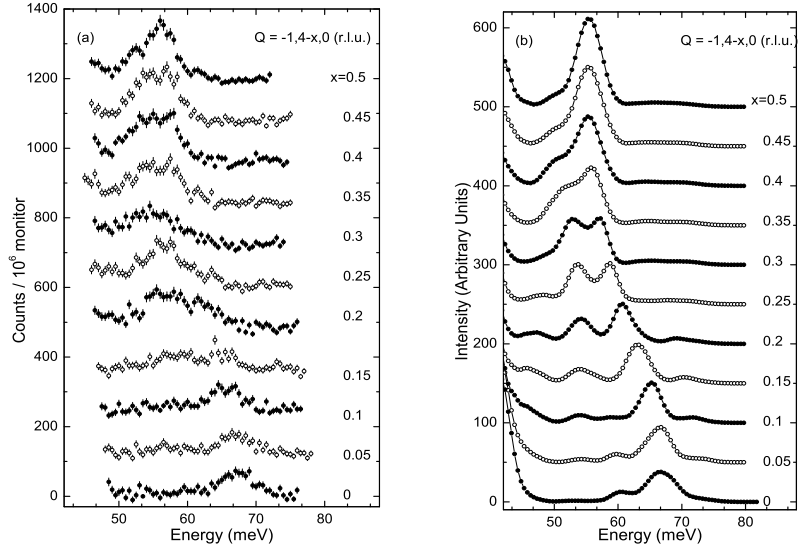}%
\caption{Measured (a) and resolution-corrected calculated (b)
neutron scattering intensity at \textbf{Q} = (-1,4-X,0) r.l.u.
Exactly the same background as in Fig. 1 has been subtracted from the experimental raw data.}%
\end{center}
\end{figure}

Such a simple model cannot reproduce the anomalous softening of
the bond-stretching phonons, and thus would not realistically
simulate the experiment. In particular, it would miss the effect
of the anomalous softening of the bond-stretching modes on other
phonons: it leads to anti-crossings with branches of the same
symmetry, which originally have a different polarization, but
hybridize with the bond-stretching branches if they are close in
energy. Special terms were added in order to model such effects:
1) A negative breathing deformability lowers the energy of the
planar breathing mode; 2) A term lowers the energies of the
'half-breathing' modes at the zone boundary of the (100) and -
with a different parameter - of the (010) direction; 3) Similar
terms produce maximum softening half way to the zone boundary in
the 100 and 010 directions, but have no effect both at the zone
center and at the zone boundary. The physical meaning of the
special terms (especially the last two) is not completely clear.
We think that they mimic the
effects of a strong electron-phonon coupling. The calculations also included a resolution correction in \textbf{Q} and $\omega$.%

The longitudinal plane oxygen vibrations with atomic displacements
in the \textbf{a} and \textbf{b} directions have the zone center
energies of 74 and 68 meV respectively (Figs. 1, 2). Their zone
boundary energies are 58 and 55 meV respectively. As the
\textbf{a}-axis polarized longitudinal phonon disperses downward
from the zone center, the peak disappears almost entirely at
\textbf{q}$_{a}$=0.15 and 0.2 and then reappears at higher q and
lower energy (55meV). Along the \textbf{b}-axis, the peak
similarly disappears and reappears somewhat closer to the zone
center. The difference between the energies of the \textbf{a} and
the \textbf{b}-axis vibrations is entirely accounted for by the
difference in the bond lengths.

\begin{figure}
[h!]
\begin{center}
\includegraphics[
width=\textwidth
]%
{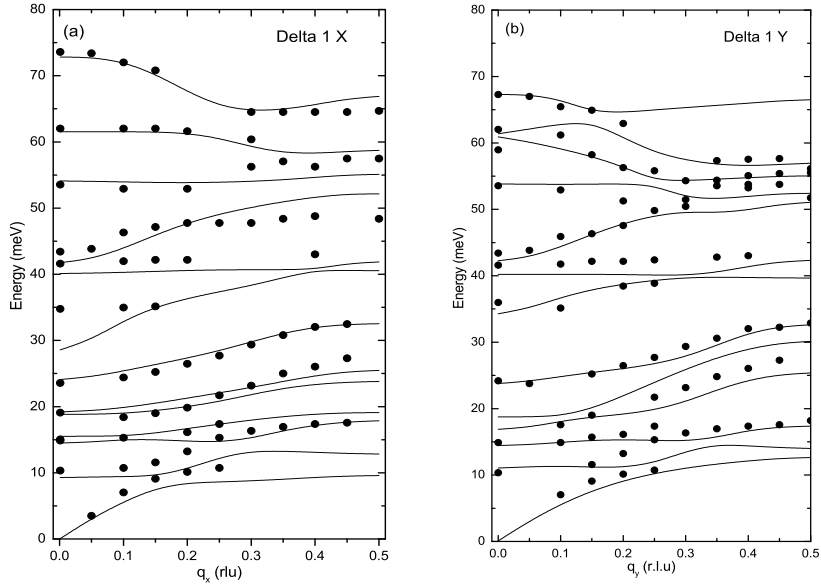}%
\caption{Observed and calculated phonon frequencies for the
$\Delta$1 branches. The lines and the dots represent calculated
and experimentally measured phonon energies for \textbf{q}
perpendicular (a) and
parallel (b) to the chains.}%
\end{center}
\end{figure}

These data seem to imply that the dispersions of the longitudinal
oxygen vibrations are discontinuous and one may be tempted to
conclude that the lattice dynamics in
YBa$_{\text{2}}$Cu$_{\text{3}}$O$_{\text{7}}$ are highly
unconventional. Similar data have been explained by unit cell
doubling or charge stripe formation\cite{3,5}.

Inspection of the figures shows that the final model reproduces
both the experimental frequencies and the observed intensities
quite well (Figs. 1b and 2b). It predicts that an almost flat
apical oxygen branch should cross the downward-dispersing
longitudinal plane oxygen branch. The apical oxygen branch does
not show up in figures 1 and 2 because of its small structure
factor in the measured Brillouin zones (in fact, zero at
$\Gamma$). Since the apical and plane oxygen branches have the
same symmetry, they do not actually cross, but mix and repel each
other lifting the degeneracy at the crossing point. This is the
branch anticrossing phenomenon described above. In the process,
the lower branch acquires plane oxygen character on the higher
\textbf{q} side of the crossing point and becomes visible in the
measured Brillouin zones. Simultaneously, the upper branch
acquires mostly apical oxygen character and becomes much weaker.%

Figure 3 shows that there is a good agreement between the
experimentally observed phonon peak positions and the calculated
dispersion curves. Furthermore, the shell model predicts the
structure factors that at least qualitatively account for
experimentally observed phonon intensities (Figs. 1, 2). Thus we
see no need to invoke unconventional theories such as charge
stripes or unit cell doubling\cite{5} to describe lattice dynamics
of YBa$_{\text{2}}$Cu$_{\text{3}}$O$_{\text{7}}$.

Finally, we found no appreciable temperature dependence of
longitudinal oxygen vibrations measured on a twinned sample other
than slight softening with increasing temperature consistent with
increasing anharmonicity.

This work was partially supported by the New Energy and Industrial
Technology Development Organization (NEDO) as Collaborative
Research and Development of Fundamental Technologies for
Superconductivity Applications. JMT is supported by the U.S.
Department of Energy's Office of Science under Contract No.
DE-AC02-98CH10886


\begin{thebibliography}{9}                                                                                                %


\bibitem {1}see for example J.M. Tranquada et al., \textit{Phys. Rev. Lett.}
\textbf{88}, 075505 (2002) and references therein.

\bibitem {2}L. Pintschovius et al., \textit{Physica C} (Amsterdam)
\textbf{185-189}, 156 (1991); W. Reichardt et al. Physica C (Amsterdam)
\textbf{162-164}, 464 (1989).

\bibitem {3}R.J. McQueeny et al., \textit{Phys. Rev. Lett.} \textbf{82}, 156 (1991).

\bibitem {4}S. L. Chaplot et al., \textit{Phys. Rev. B} \textbf{52}, 7230 (1995).

\bibitem {5}R.J. McQueeny et al., cond-mat/0105593
\end{thebibliography}
\end{document}